%% file: main.tex
\documentclass[format=acmsmall, nonacm, screen=True]{acmart}
\usepackage{tikz}
\usepackage{threeparttable}
\usepackage{array}
\usepackage{geometry}
\usepackage[utf8]{inputenc}
\usepackage{graphicx} 
\usepackage{pgfplots}
\usepackage{caption}
\usepackage{graphicx}
\usepackage{bbm}
\usepackage{threeparttable}
\usepackage{soul}
\usepackage[misc]{ifsym}
\usepackage[tight]{subfigure}
\usepackage[capitalize]{cleveref}
\usepackage{rotating}
\usepackage{float}
\usepackage{longtable}
\usepackage{array}
\usepackage{amsmath}
\usepackage{tabularx}
\usepackage{ragged2e}
\usepackage{lipsum}
\usepackage{longtable}
\usepackage{enumitem}
\usepackage{setspace}
\usepackage{balance}
\pgfplotsset{compat=1.17}

\usepackage{makecell}

\usepackage[subtle,tracking=normal, wordspacing=normal,mathdisplays=tight]{savetrees}
\newif\ifshowcomment
\showcommenttrue
\ifshowcomment
\newcommand{\luyao}[1]{\textcolor{blue}{[luyao] #1}}

\else
\newcommand{\luyao}[1]{}
\fi

\setcitestyle{authoryear}

\begin{document}

\title{WIP: The Design Principle of blockchain: \\ An Initiative for the SoK of SoKs}


\author{Luyao Sunshine Zhang}
\affiliation{%
  \institution{Primitives Lane}
  \country{Metaverse}}
  \authornote{Primitives Lane is a nonprofit research group focused on blockchain and other frontier technologies. We are dedicated to solving the most fundamental public issues in frontier fields, helping researchers grow steadily, and creating a friendly and supportive space for builders.}

\renewcommand{\shortauthors}{Sunshine Zhang}

\begin{abstract}
\begin{quote}
The Master said, "\textbf{virtue} is not left to stand alone. He who practices it will have neighbors."\\---Confucius, \textit{Analects}
\end{quote}

\begin{quote}
“You must contrive for your future rulers another and a better life than that of a ruler, and then you may have a well-ordered State; for only in the State which offers this, will they rule who are truly rich, not in silver and gold, but in \textbf{virtue} and wisdom, which are the true blessings of life.”\\--- Plato, \textit{Republic}
\end{quote}

Blockchain, also coined as decentralized AI, has the potential to empower AI to be more trustworthy by creating a decentralized trust of privacy, security, and audibility. However, systematic studies on the design principle of blockchain as a trust engine for an integrated society of cyber-physical-social-system (CPSS) are still absent. In this article, we provide an initiative for seeking the design principle of blockchain for a better digital world. Using a hybrid method of qualitative and quantitative studies, we examine the past origin, the current development, and the future directions of blockchain design principles. We have three findings. First, the answer to whether blockchain lives up to its original design principle as a distributed database is controversial. Second, the current development of the blockchain community reveals a taxonomy of 7 categories, namely, privacy and security, scalability, decentralization, applicability, governance and regulation, system design, and cross-chain interoperability. Both research and practice are more centered around the first category of privacy and security and the fourth category of applicability. Future scholars, practitioners, and policy-makers have vast opportunities in other, much less exploited facets and the synthesis at the interface of multiple aspects. Finally, in counter-examples, we conclude that a synthetic solution that crosses discipline boundaries is necessary to close the gaps between the current design of blockchain and the design principle of a trust engine for a truly intelligent world.  

\textbf{Acknowledgments:} I am deeply indebted to all the pioneers in Web3 for their inspirations, especially for my mentors, coauthors, and students who jointly contribute to the interdisciplinary conversation around the applications of blockchain. \textit{The name list is to be added.} 
\end{abstract}

\ccsdesc[500]{Applied computing~Economics}
\ccsdesc[500]{Security and privacy~Distributed systems security}
\ccsdesc[500]{Human-centered computing~Collaborative and social computing systems and tools}

\keywords{blockchain, AI ethics, philosophy, economics, computer science}

\begin{teaserfigure}
  \includegraphics[width=\textwidth]{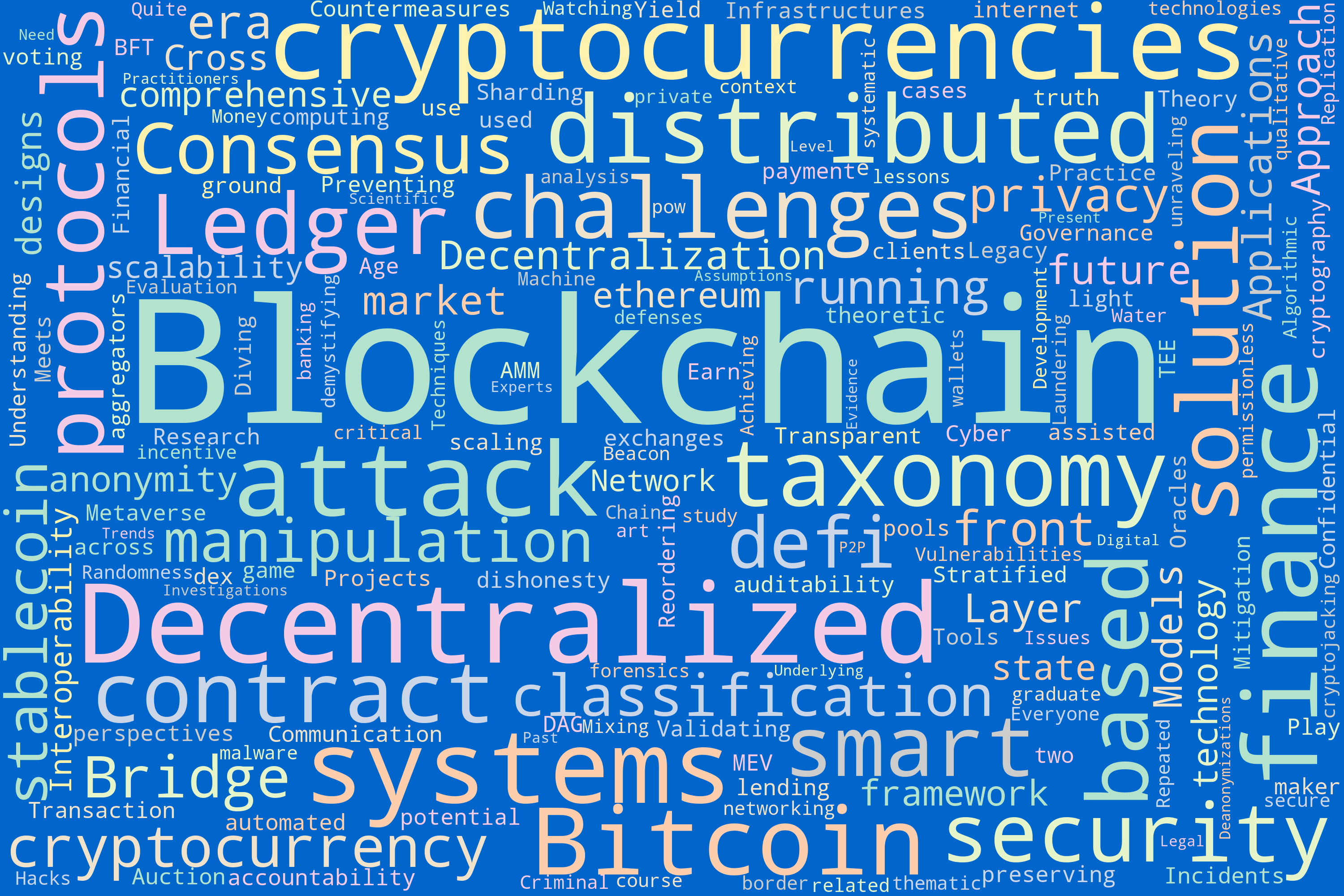}
  \caption{The Word Cloud of blockchain Related SoK Titles}
  \label{fig:teaser}
\end{teaserfigure}

\maketitle

\section{Introduction}
The past one hundred years have witnessed incredible advancements in artificial intelligence (AI)~\cite{stone2022artificial,littman2022gathering}. These advancements are integrating cyberspace (CS), physical space (PS), and social space(SS), into the cyber-physical-social-system(CPSS), which extraordinarily expands the territories of human civilizations~\cite{wang2010emergence} extraordinarily. However, AI \textit{per se} is not enough to establish trust in the CPSS~\cite{jacovi2021formalizing,jan2020ai}, which is the cornerstone of prosperity in every civilized society. Blockchain, also coined as \textit{decentralized AI}, has the potential to empower AI to be more trustworthy by creating a decentralized trust of privacy, security, and audit-ability~\cite{harris2019decentralized,adel2022decentralizing,hussain2021artificial}. However, systematic studies on the design principle of blockchain as a trust engine for the CPSS are still absent. If we could decipher the philosophy of blockchain, we would build the infrastructure of a better digital world. In this article, we provide an initiative for seeking the design principle of blockchain for the betterment of human civilization. Unlike the existing systemization of knowledge (SoK) on blockchain that focuses on a specific discipline purpose or topic of interest, ours aims at opening an intellectual conversation beyond boundaries for the ultimate goal of a better digital world, namely, the SoK of SoKs. Specifically, we hope to initiate the answers to three questions for the past, the current, and the future of blockchain design.
\begin{enumerate}
    \item {\em The past:} Does blockchain live up to its original design principles as a distributed database? 
    \item {\em The current:} How do the current blockchain literature, industry practices, and global standards address and develop the design principle of blockchain?
    \item {\em The future:} What are the gaps between the current design of blockchain and the design principle of a trust engine for a truly intelligent world?
\end{enumerate}

In \Cref{sec: the past}, we examine the performance of blockchain according to its original design principle as a distributed database. We present the controversial answers in a dialog style of debate. In \Cref{sec: the current}, we analyze the current research and practice of blockchain principles by investigating the current SoKs of emerging blockchain literature, the white papers that provide technique cornerstones for blockchain technology in the real world, and the discussions on global standards for blockchain. We identify a taxonomy of blockchain literature into the seven categories of privacy and security, scalability, decentralization, applicability, governance and regulation, system design, and cross-chain interoperability. Integrating the AI of natural language processing (NLP)~\cite{bird2009natural} methods, we find that the current blockchain design principles, in both research and practice, are more centered around the first category of privacy and security and the fourth category of applicability. Future scholars, practitioners, and policy-makers have vast opportunities in the other much less exploited categories and the synthesis at the interface of multiple categories. In \Cref{sec: the future}, we envision the future of blockchain technology by pointing out the gaps between the current design of blockchain and the design principle of a trust engine for a truly intelligent world. By providing counterexamples, we question the possibility of developing a plausible solution of singularity to the gaps without crossing the current boundaries of domain expertise.
\par
Sages in both the West and East have been craving an ideal society. In the \textit{Book of Rites}, one of the Confucian (551–479 B.C.E.) classics~\cite{sep-confucius}, the \textit{Great Unity} is a Chinese vision of an ideal world in which men of virtue and ability rule. In ancient Greece, Plato~\cite{sep-plato} (429–347 B.C.E.) envisioned an ideal city-state ruled by a philosopher-king of justice in the \textit{Republic}, a Socratic dialog. By such a coincidence, both Confucius and Plato place the foundation of an ideal society on cultivating or selecting the men of virtue to be the administrators whom people can trust. Nevertheless, the existence of a benevolent ruler might not be guaranteed. In contrast, blockchain enables a new way of trust creation empowered by computer intelligence beyond human ethics. However, the current blockchain technology is evidently not a panacea for doubts about human trust. Then, why "throw the baby out with the bathwater?" Existing mechanisms for cultivating human trust have created economic prosperity and human civilizations for thousands of years. How then can we integrate blockchain technology into the existing governance, economics, and ethics to create an ideal society in the new era of CPSS? We initiate an open dialog of synthetic solutions. 
\section{The past}
\label{sec: the past}

\begin{figure}
  \includegraphics[width=\textwidth]{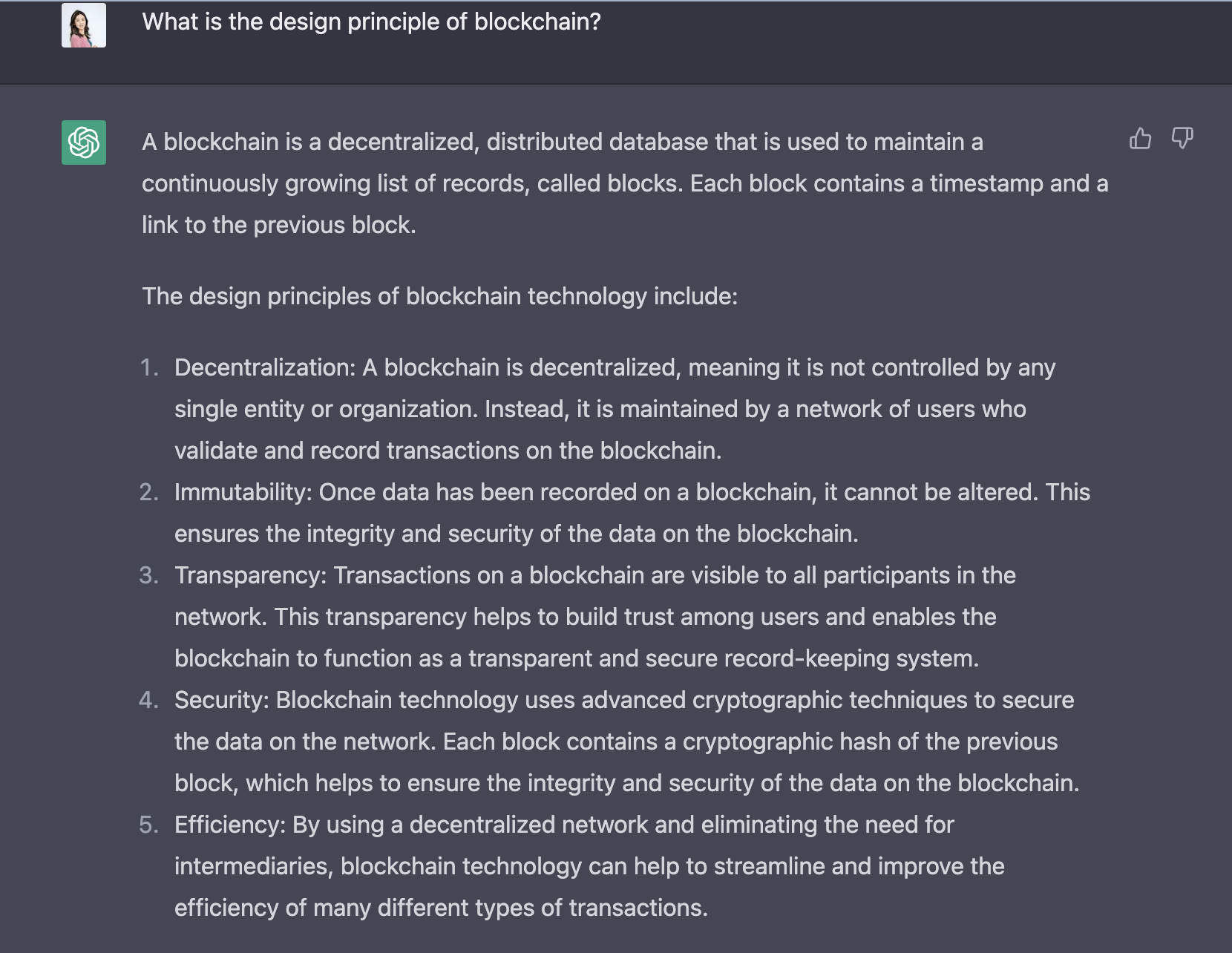}
  \caption{A Conversation Between Sunshine and Chat GPT-3 on 2022/12/30}
  \label{fig:chatGPT3}
\end{figure}

What was blockchain originally designed for? How does blockchain live up to the design principle of its original purpose? \citet{sherman2019origins} introduce the origin of blockchain for establishing a distributed database trusted by mutually suspicious groups. \citet{narayanan2016bitcoin, haber1990time},and~\citet{bayer1993improving} further elaborate on the follow-up development of blockchain technology for a series of desired properties such as decentralization, immutability, transparency, security, and efficiency. \Cref{fig:chatGPT3} shows the answer returned by the Generative Pre-trained Transformer (ChatGPT),~\url{https://openai.com/blog/chatgpt/}, a chatbot launched by OpenAI in November 2022, to the question of "what is the design principle of blockchain?" The answer accurately matches elaborations provided in earlier blockchain literature. We now examine the performance of blockchain on each principle in a dialog style of debate in the spirit of Confucius's \textit{the Analects}~\cite{ames2010analects} and Plato's \textit{the Republic}~\cite{plato2005republic}. We find the answer to be controversial. Moreover, the desired principles might either conflict with each other or lead to other undesired consequences.  
\subsection*{Question 1: Does blockchain live up to its promise of decentralization?}
\subsubsection{Pros}
Yes, blockchain is decentralized. Compared with centralized ledgers, blockchain is not subject to the single point failure of a centralized entity~\cite{puthal2018blockchain}. Blockchain records and validates data by a consensus protocol involving decentralized entities, which has the fault tolerance of minorities~\cite{zhang2022blockchain}. 
\subsubsection{Cons}
No, blockchain is not decentralized. \citet{sai2021taxonomy} provide a taxonomy of blockchain centralization in 13 aspects. For example, \citet{sai2021taxonomy} show that the top four mining pools in Ethereum (Bitcoin) consist of 63.04\% (50.36\%) of the hashing power, which is enough to conduct a successful malicious attack. Moreover, \citet{zhang2022sok} and the references therein show that the usage of blockchain at the application layer has the trend of conversing to centralization.   

\subsection*{Question 2: Does blockchain live up to its promise of immutability?}
\subsubsection{Pros}
Yes, blockchain is immutable. ~\citet{narayanan2016bitcoin} elaborate on the difficulty of manipulating recorded data due to the chain structure of Merkle trees. 
\subsubsection{Cons}
No, blockchain is mutable. \citet{hofmann2017immutability} show that the immutability of blockchain can be breached by various attacks. \citet{politou2019blockchain} even address the conflicts between blockchain immutability and the new European data protection regulation on the right to be forgotten (RtbF), according to which individuals have the right to delete their personal data under certain conditions. 

\subsection*{Question 3: Does blockchain live up to its promise of transparency?}

\subsubsection{Pros}
Yes, blockchain is transparent. ~\cite{monrat2019survey} explains that all the transactions on blockchain are recorded and auditable by each replica node in the network. 
\subsubsection{Cons}
No, blockchain is not transparent. \citet{sai2021taxonomy} demonstrates the centralization in node operation, which is required to audit the blockchain data due to technique barriers. \citet{feng2019survey} even point out the transparency of blockchain as a negative feature and a threat to user privacy. 

\subsection*{Question 4: Does blockchain live up to its promise of security?}
\subsubsection{Pros}
Yes, blockchain is secure. \citet{zhang2019security} address the blockchain security property of consistency and integrity achieved by the main techniques of consensus algorithms and hash-chained storage. 
\subsubsection{Cons}
No, blockchain is not secure. \citet{lin2017survey} evidence a long list of real attacks on blockchain together with an analysis of potential security issues. \citet{budish2022economic} even establishes a conclusion that the mechanism supporting blockchain security is either vulnerable or not worth the economic cost. 

\subsection*{Question 5: Does blockchain live up to its promise of efficiency?}
\subsubsection{Pros}
Yes, blockchain is efficient. \citet{wang2019blockchain} analyze how blockchain improves efficiency in human interactions by automating and streamlining business operations using smart contracts. 
\subsubsection{Cons}
No, blockchain is not efficient. Current major blockchains can only sustain tens of transactions per second (TPS), which is not comparable to the centralized platforms, which have a throughput of thousands of TPS~\cite{chauhan2018blockchain}.

\section{The current}
\label{sec: the current}
How do the current blockchain literature, industry practices, and global standards address and develop the design principle of blockchain? We collect all the systemization of knowledge (SoK) papers related to facets of blockchain or surveys and review articles that satisfy the defining features of SoKs. We refer the readers to the JSys website, \url{https://www.jsys.org/type_SoK/}, for background information on SoKs and the Oakland Conference GitHub site, \url{https://oaklandsok.github.io/}, for the histories of SoKs in computer security and privacy literature. We present a taxonomy of the collected SoKs in \Cref{tab1} and \Cref{tab2}, broken into seven categories based on the topic of research questions: 
\begin{itemize}
    \item {\em Category 1: Privacy and Security.} We include literature that answers questions about private information protection and attacks on blockchain systems. We refer the readers to \citet{sep-privacy} for a philosophical symposium on \textit{privacy} and \citet{bishop2003computer} for the computer science insights on \textit{security}.
    \item {\em Category 2: Scalability.} We include literature that answers questions related to the scalable deployment of blockchain, usually measured in throughput and quality of service. We refer the readers to \citet{jogalekar2000evaluating} for the definition and measurement of scalability in the distributed system literature. 
    \item {\em Category 3: Decentralization.} We include the literature that answers questions related to the designed and realized decentralization of blockchain. We refer the readers to \citet{zhang2022sok} for an interdisciplinary synthesis of a taxonomy of blockchain decentralization. 
    \item {\em Category 4: Applicability.} We include literature that answers questions related to the application of blockchain to solve social and economic issues in various of human interactions. \Cref{tab2} shows that the major applications are in financial services~\cite{zetzsche2020decentralized,chen2020blockchain,harvey2021defi}. 
    \item {\em Category 5: Governance and Regulation.} We include literature that answers questions related to utilizing blockchain as a new way of governance, the process of interaction of an organized society, and its interactions with existing governance solutions, including corporate, government, and nongovernment organizations (NGOs). We refer the readers to \cite{bevir2012governance} for a review of existing governance solutions, including various forms of social coordination and patterns of rules.    
    \item {\em Category 6: System design.} We include literature that answers questions related to designing blockchain as an advancement of existing computer systems. We refer the readers to \citet{saltzer2009principles} for the design principles of computer systems. 
    \item {\em Category 7: Cross-chain and Interoperability.} We include literature that answers questions related to communication, cooperation, and integration among blockchain systems, other computer systems, and human societies. We refer the readers to \citet{wegner1996interoperability} and \citet{leal2019interoperability} for the importance and assessment of interoperability among information systems. 
\end{itemize}

\Cref{tab6} lists the top 28 blockchain projects, excluding applications on the blockchain system and cross-chain solutions ranked by the market value of their native currency retrieved from Coinmarketcap, \url{https://coinmarketcap.com/}, on Dec. 27, 2022. We further collect the white papers that provide the technical foundations for the 28 projects. We then conduct a text analysis applying natural language processing (NLP) methods to produce a word cloud and bigram networks of the titles and abstracts of the SoKs and white papers. The results are presented in \Cref{tab4}, \Cref{tab5}, \Cref{tab7}, \Cref{tab8}, \Cref{tab9}, \Cref{fig:teaser}, \Cref{fig:whitepaper_bigram_title}, \Cref{fig:wordcloud_abstract}, \Cref{fig:whitepaper_wordcloud_abstract}, \Cref{fig:whitepaper_wordcloud_title}, \Cref{fig:bigram_title}, \Cref{fig:bigram_abstract}, and \Cref{fig:whitepaper_bigram_abstract}, \Cref{fig:whitepaper_bigram_title}. The word cloud distinguishes the word frequency in these documents through font size, and the bigram represents the coappearance of two words in sequential order ranking by counts in the tables and in the network figures. We also further research the emerging documents of blockchain standard development and papers discussing blockchain standards by working groups and institutions globally.  
\par
\textbf{Data and Code Availability}: We open source the data and code for replication and future research on the GitHub: \url{https://github.com/sunshineluyao/design-principle-blockchain}. 
    
We find that the current blockchain design principles, in both research and practice, are more centered around the first category of privacy and security and the fourth category of applicability. Future scholars, practitioners, and policy-makers have vast opportunities in the other much less exploited categories and the synthesis at the interface of multiple categories.
\section{The future}
\label{sec: the future}
What are the gaps between the current design of blockchain and the design principle of a trust engine for a truly intelligent world? In this section, we do not intend to provide a comprehensive answer. Instead, we question the possibility of developing a plausible solution of singularity to the gaps without crossing the current boundaries of domain expertise by providing counterexamples.

\subsection*{Case Study 1: \\ How can we elaborate on the design principle of privacy and security for a better society?}
The scientific community of distributed systems and cryptography has made great milestones in blockchain development by achieving the anonymity of private information and the audibility of public transactions. However, privacy and security are the means but not necessarily the ends for a better society of human prosperity. Through history, critiques~\cite{sep-privacy} have addressed privacy as the source of criminal activities, economic inefficiency, and the abuse of minorities. \citet{cong2022crypto} identify that wash trading accounts for approximately 70\% of the total cryptocurrency trading volumes. \citet{cong2022anatomy} further provide a taxonomy of crimes enabled by the anonymity feature on blockchain. Abundant literature in behavioral science~\cite{dawson2018study} establishes a connection between anonymity and abusive behavior. However, none of those negative effects on human behavior are considered in the original design principles of blockchain. How can we redesign privacy and security to cultivate human cooperation better and minimize abusive behavior? Behavior scientists have the expertise to contribute to this discussion. 

\subsection*{Case Study 2: \\ How can we elaborate on the design principle of decentralization for a better society?}
Most top-ranked blockchain projects aim to create a permissionless system that is more inclusive, democratic, and decentralized than those that have been presented previously. However, what is required to satisfy the permission for the permissionless blockchains? You must have access to the internet, afford the computer system requirements to run a replica node, and master the engineering skills needed to participate in the network. Unfortunately, according to World in Data, \url{https://ourworldindata.org/internet}, in most parts of our world, less than half of the population has access to the internet, let alone the higher requirements of computer system and engineering skills. Moreover, \citet{ao2022decentralized} and the references therein evidence most of the crypto transactions are executed via centralized exchanges but not in the decentralized form of on-chain peer-to-peer transactions. \citet{cong2022inclusion} further address the inclusion issues of the current Web3 economy due to technology barriers and lack of inclusion consideration in the original design principle of blockchain. The famous Irish playwright Bernard Shaw said, "Revolutions have never lightened the burden of tyranny. They have only shifted it to another shoulder"~\cite{tiliouine2016state}. To direct blockchain development to avoid the fate stated by Bernard Shaw and serve a more decentralized society---of the people, by the people, and for the people---anthropologists, social scientists, and economists are able to contribute to designing a society of democracy, diversity, and inclusion. 

\subsection*{Case Study 3: \\ How can we elaborate on the design principle of efficiency for a better society?}

The current solutions for efficiency or scalability on blockchain focus on improving transaction throughput and automating business processes. However, the pursuit of efficiency in an isolated system of short-sighted consideration might lead to inefficiency for the society as a whole with regard to long-term sustainability. For example, the current scalability solutions in general consider the blockchain system only in isolation but ignore the negative externality of the high-energy consumption design to the outside world~\cite{truby2018decarbonizing}. Moreover, \citet{grimmelmann2019all} notes that although smart contracts automate business processes, the current design ignores the ambiguity in the semantics of the programming language. Thus, a temporary convenience of the smart contract operation might lead to costly disputes in unforeseen scenarios when developers and users disagree. Policy-makers and lawyers can contribute to reconsidering the blockchain design to improve the efficiency of negotiations among different stakeholders and our society as a whole.

\subsection*{A Call for Collaboration}
\citet{kranzberg1986technology} stated the following in his 1985 address as president of the Society for the History of Technology (SHOT): "Technology is neither good nor bad, nor is it neutral." blockchain is a powerful technology that brings about many new possibilities. How can we redesign blockchain for a truly intelligent world? To address this questions, we call for a joint endeavor from all disciplines. 

\bibliographystyle{ACM-Reference-Format}
\bibliography{cite}
\appendix
\input{tabs/tab1.tex}
\input{tabs/tab2.tex}
\input{tabs/tab3.tex}
\input{tabs/tab4.tex}
\input{tabs/tab5.tex}
\input{tabs/tab6.tex}
\input{tabs/tab7.tex}
\input{tabs/tab8.tex}
\input{tabs/tab9.tex}

\begin{figure}
  \includegraphics[width=\textwidth]{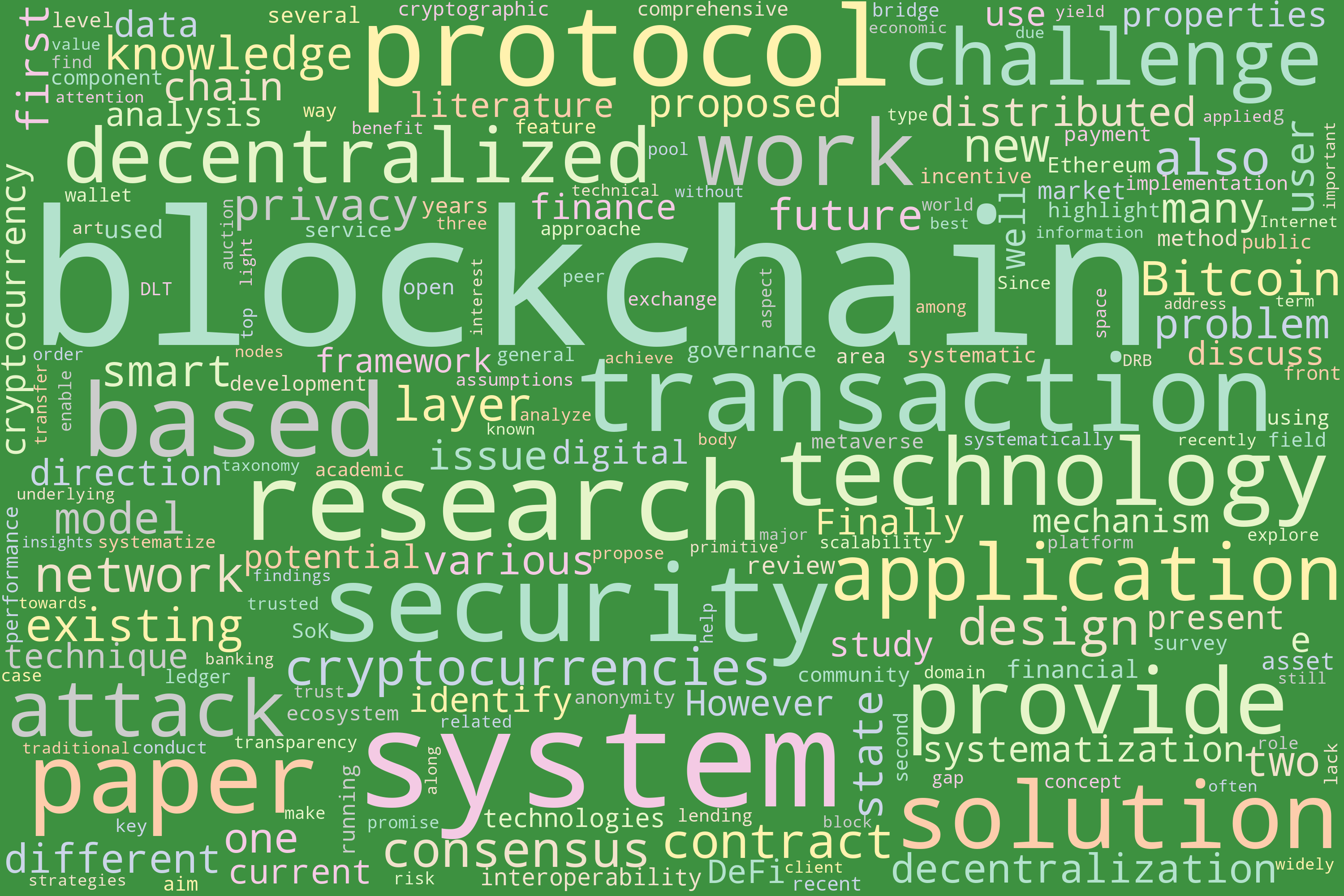}
  \caption{The Word Cloud of blockchain Related SoK Abstract}
  \label{fig:wordcloud_abstract}
\end{figure}

\begin{figure}
  \includegraphics[width=\textwidth]{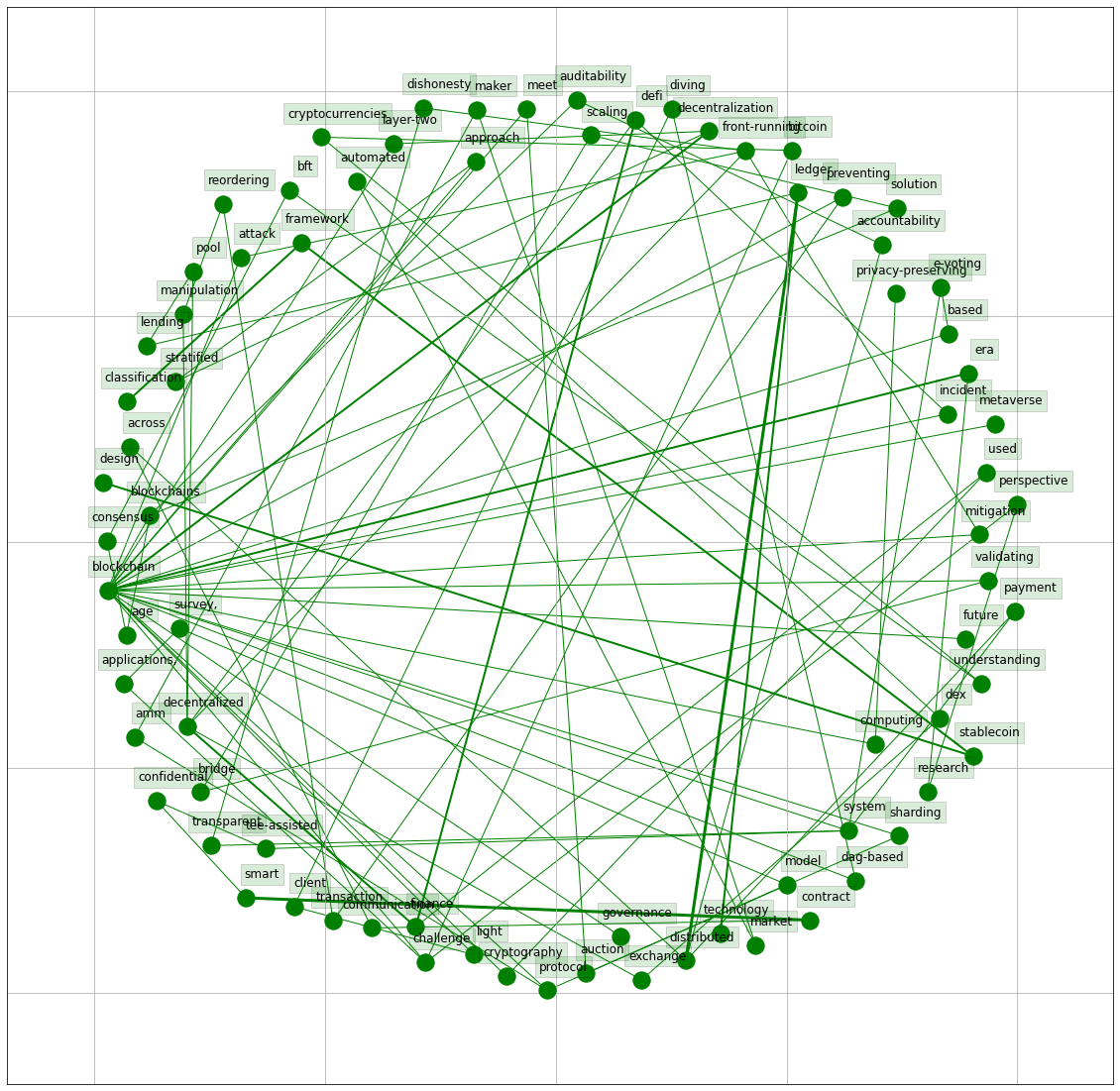}
  \caption{The Bigram of blockchain Related SoK Title}
  \label{fig:bigram_title}
\end{figure}

\begin{figure}
  \includegraphics[width=\textwidth]{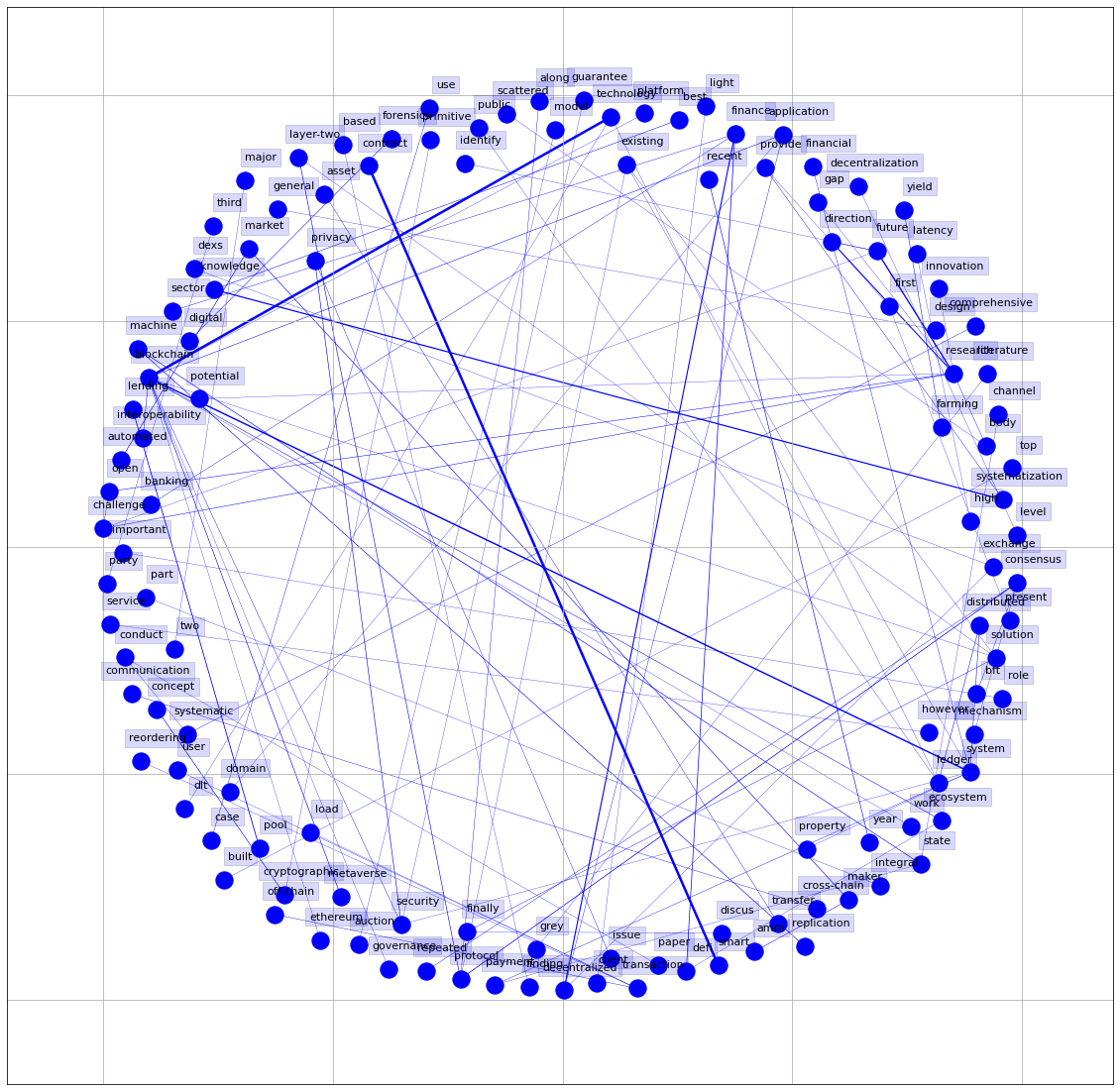}
  \caption{The Bigram of blockchain Related SoK Abstract}
  \label{fig:bigram_abstract}
\end{figure}

\begin{figure}
  \includegraphics[width=\textwidth]{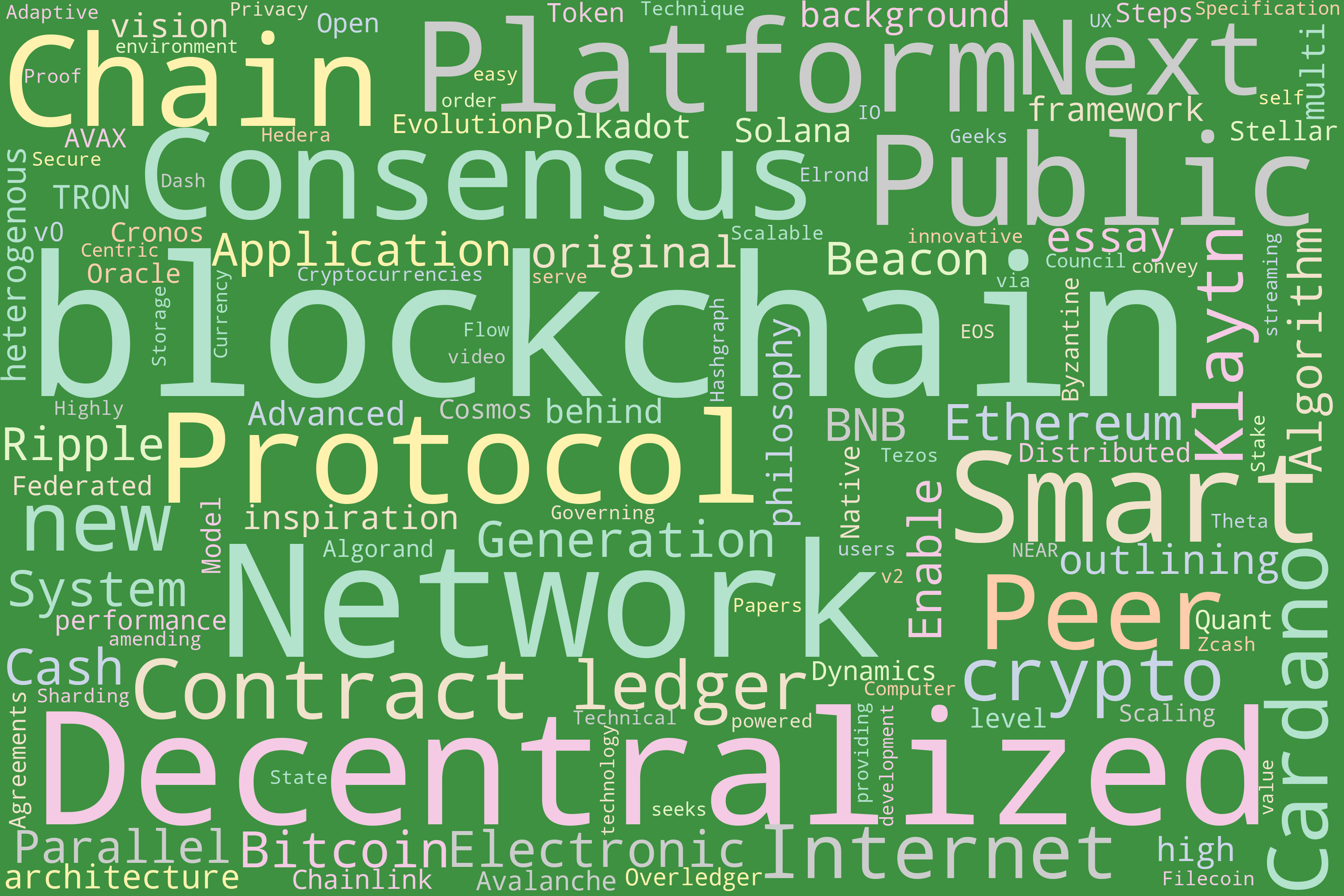}
  \caption{The Word Cloud of blockchain Projects and Cross-chain Solutions Title}
  \label{fig:whitepaper_wordcloud_title}
\end{figure}
\begin{figure}
  \includegraphics[width=\textwidth]{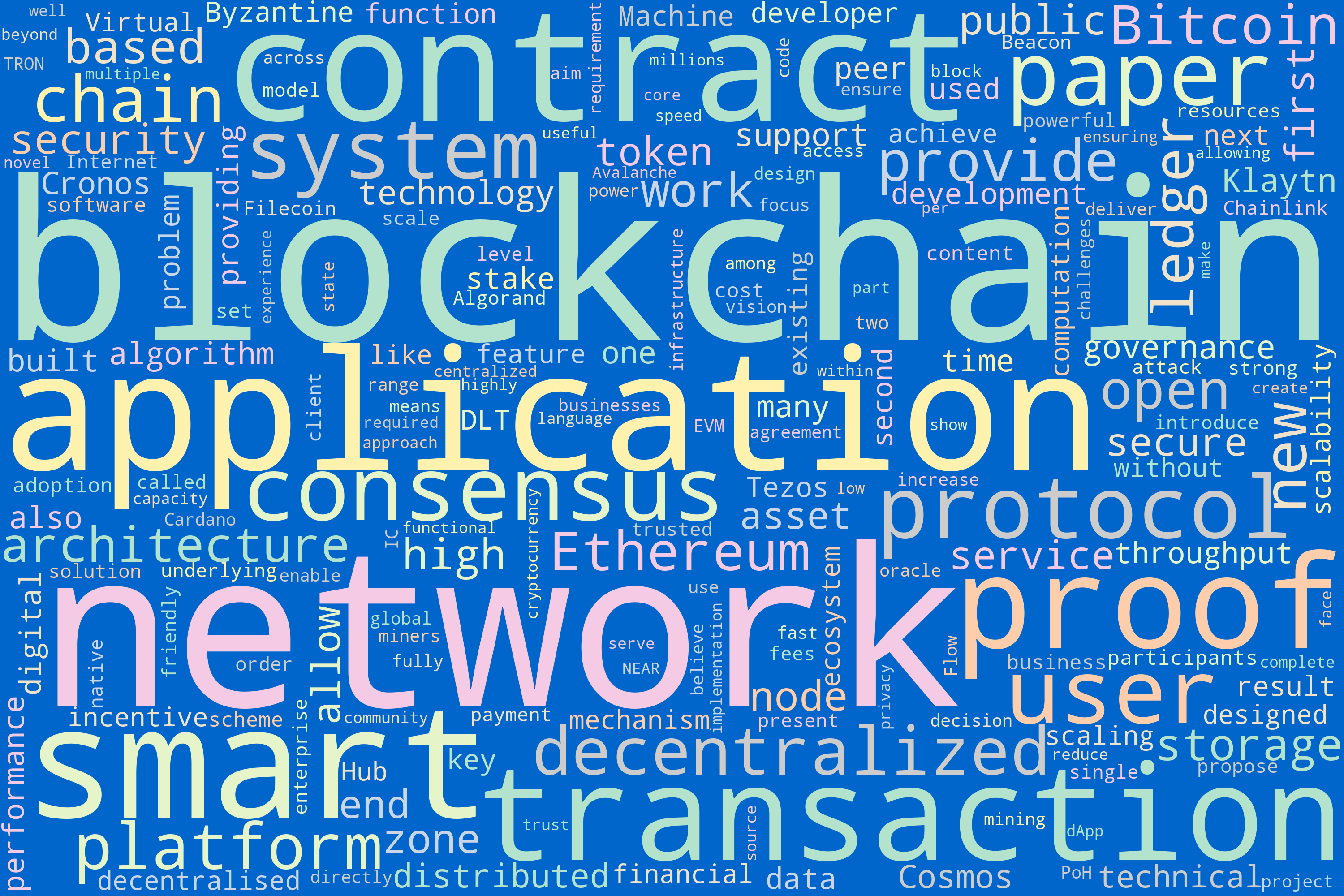}
  \caption{The Word Cloud of blockchain Projects and Cross-chain Solution Abstract}
  \label{fig:whitepaper_wordcloud_abstract}
\end{figure}

\begin{figure}
  \includegraphics[width=\textwidth]{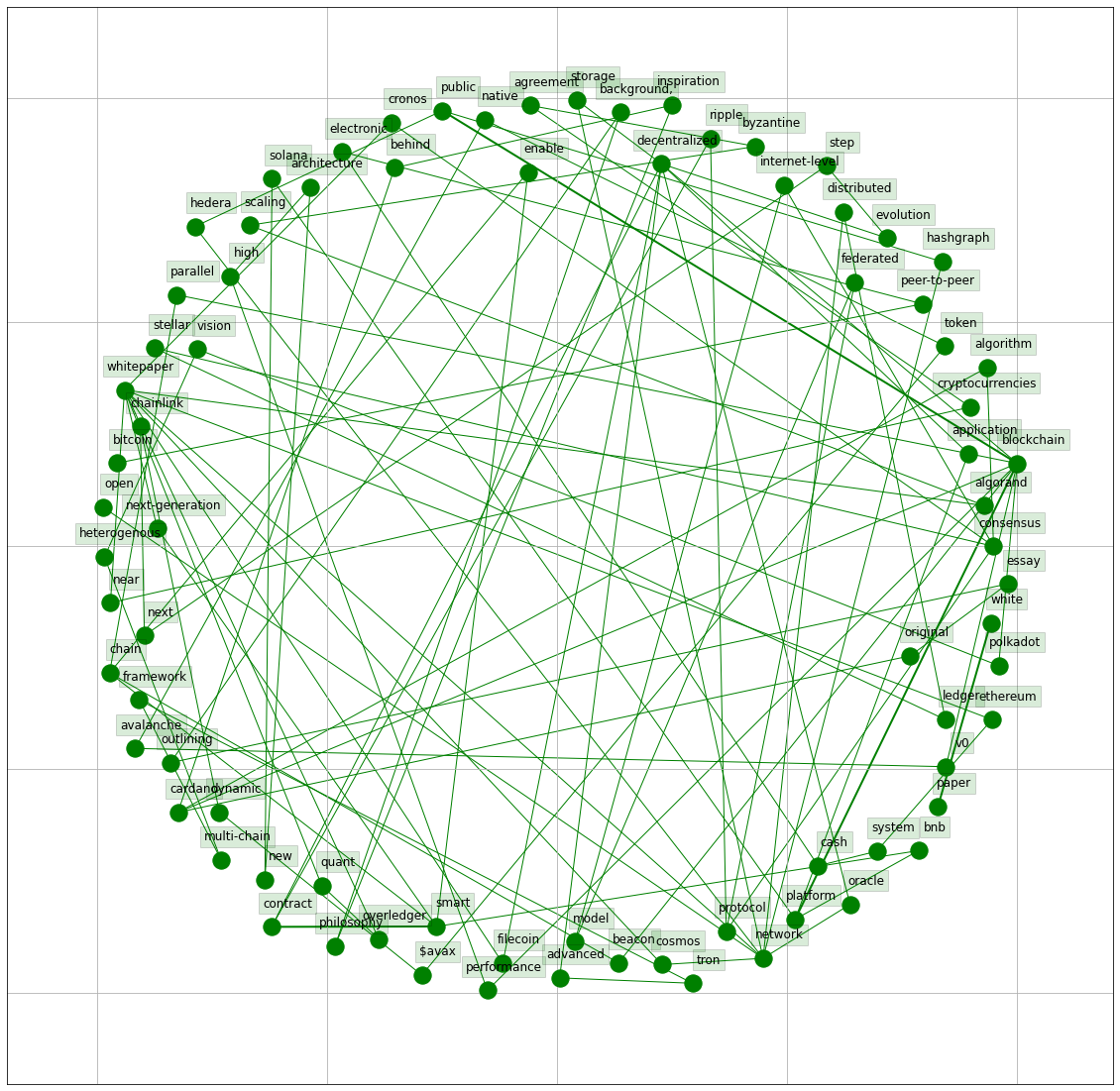}
  \caption{The Bigram of blockchain Projects and Cross-chain Solution Title}
  \label{fig:whitepaper_bigram_title}
\end{figure}

\begin{figure}
  \includegraphics[width=\textwidth]{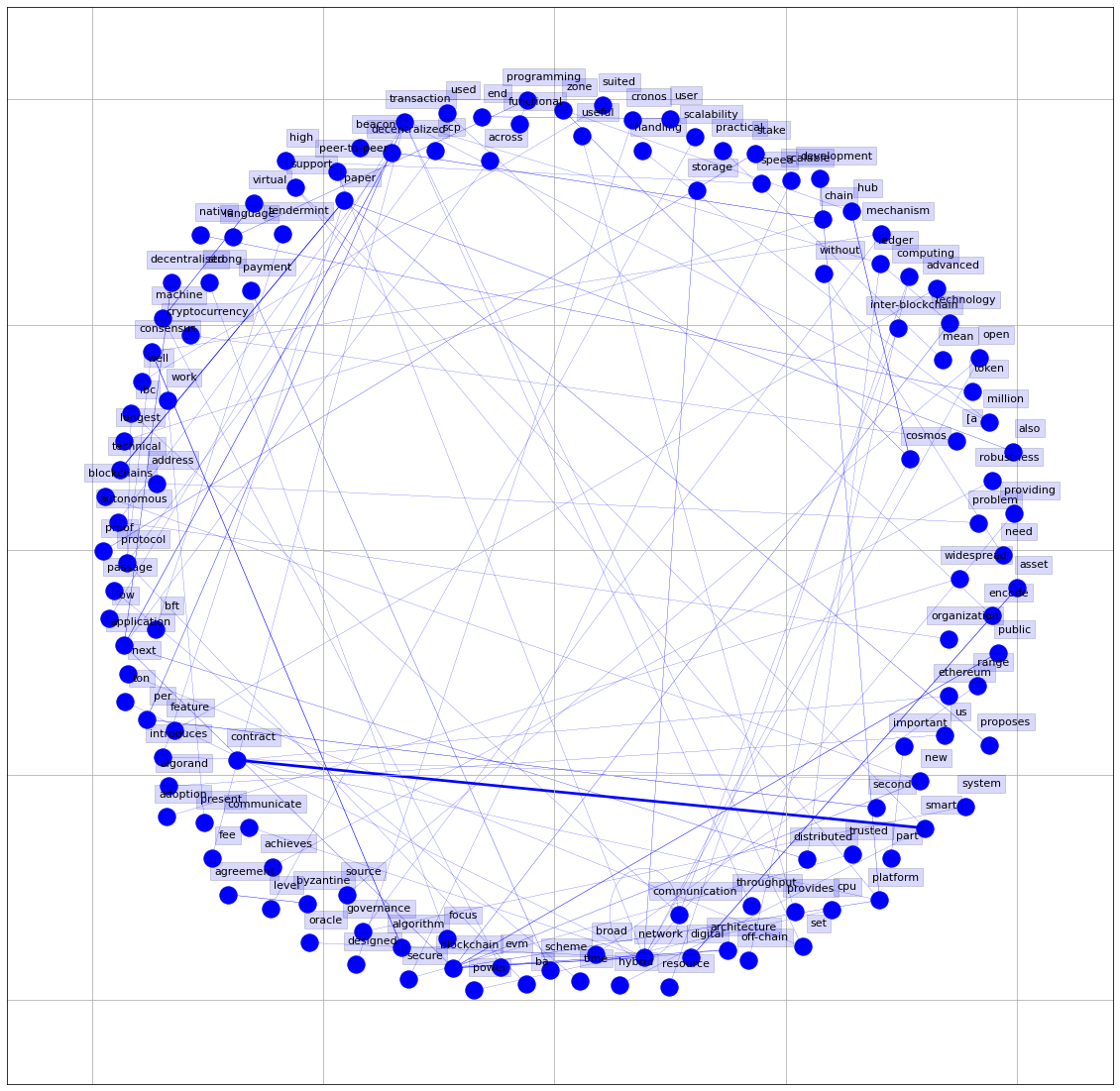}
  \caption{The Bigram of blockchain Projets and Cross-chain Solution Abstract}
  \label{fig:whitepaper_bigram_abstract}
\end{figure}

\end{document}

%% file: tabs/tab1.tex
\begin{table}[!htbp]
\fontsize{7pt}{\baselineskip}\selectfont
\setlength{\tabcolsep}{4pt}
\begin{tabular}{|l|r|}
 \hline \hline
Citation & Title \\ [1ex] 
 \hline \hline
 \multicolumn{2}{r}{\textbf{Category 1: Privacy and Security}} \\ [1ex] 
 \hline
\citet{almashaqbeh2022sok} & SoK: privacy-preserving computing in the blockchain era\\[0.2ex]  
 \hline
 \citet{bonneau2015sok} & SoK: Research Perspectives and Challenges for Bitcoin and Cryptocurrencies \\[0.2ex] 
\hline
 \citet{wang2021sok} &  SoK: Understanding BFT Consensus in the Age of Blockchains \\[0.2ex] 
\hline
 \citet{eskandari2020sok} &  Sok: Transparent dishonesty: front-running attacks on blockchain \\[0.2ex] 
\hline
 \citet{baum2021sok} &  Sok: Mitigation of front-running in decentralized finance \\[0.2ex] 
\hline
 \citet{raikwar2019sok} &  SoK of used cryptography in blockchain \\[0.2ex] 
\hline
 \citet{heimbach2022sok} &  SoK: Preventing Transaction Reordering Manipulations in Decentralized Finance \\[0.2ex] 
\hline
\citet{zhou2022sok} &  SoK: Decentralized Finance (DeFi) Incidents \\[0.2ex] 
\hline
\citet{li2022sok} &  SoK: TEE-assisted Confidential Smart Contract \\[0.2ex] 
\hline

\citet{ankele2020sok} &  SoK: Cyber-Attack Taxonomy of Distributed Ledger-and Legacy Systems-based Financial Infrastructures \\[0.2ex] 
\hline

\citet{yang2022sok} &  SoK: MEV Countermeasures: Theory and Practice \\[0.2ex] 
\hline

\citet{azouvi2019sok} &  Sok: Tools for game theoretic models of security for cryptocurrencies \\[0.2ex] 
\hline

\citet{atzei2017survey} &  A survey of attacks on ethereum smart contracts (sok) \\[0.2ex] 
\hline

\citet{judmayer2021sok} &  Sok: Algorithmic incentive manipulation attacks on permissionless pow cryptocurrencies \\[0.2ex] 
\hline

\citet{di2020sok} &  SoK: Development of secure smart contracts--lessons from a graduate course \\[0.2ex] 
\hline

\citet{chen2020survey} &  A survey on ethereum systems security: Vulnerabilities, attacks, and defenses \\[0.2ex] 
\hline

\citet{islam2021review} &  A Review on Blockchain Security Issues and Challenges \\[0.2ex] 
\hline

\citet{li2020survey} &  A survey on the security of blockchain systems \\[0.2ex] 
\hline

\citet{garay2020sok} & Sok: A consensus taxonomy in the blockchain era \\[0.2ex] 
\hline

\citet{tekiner2021sok} & SoK: cryptojacking malware \\[0.2ex] 
\hline

\citet{alsalami2019sok} & SoK: A systematic study of anonymity in cryptocurrencies \\[0.2ex] 
\hline

\citet{deuber2022sok} & SoK: Assumptions Underlying Cryptocurrency Deanonymizations \\[0.2ex] 
\hline

\citet{rinberg2022privacy} & Privacy when Everyone is Watching: An SOK on Anonymity on the Blockchain \\[0.2ex] 
\hline

\citet{bonomi2021sok} & SoK: Achieving State Machine Replication in Blockchains based on Repeated Consensus \\[0.2ex] 
\hline

\citet{ghesmati2021sok} & SoK: How private is Bitcoin? Classification and Evaluation of Bitcoin Mixing Techniques \\[0.2ex] 
\hline

\citet{franzoni2022sok} & SoK: Network-Level Attacks on the Bitcoin P2P Network \\[0.2ex] 
\hline


\multicolumn{2}{r}{\textbf{Category 2: Scalability}} \\ [1ex] 
 \hline
\citet{wang2019sok} &  SoK: Sharding on Blockchain\\[0.2ex] 
\hline
\citet{gudgeon2020sok} &  SoK: Layer-Two Blockchain Protocols \\[0.2ex] 
\hline

\citet{mccorry2021sok} &  Sok: Validating bridges as a scaling solution for blockchains \\[0.2ex] 
\hline

\citet{chatzigiannis2022sok} &  Sok: Blockchain light clients \\[0.2ex] 
\hline

\citet{zhou2020solutions} &  Solutions to scalability of blockchain: A survey \\[0.2ex] 
\hline

\multicolumn{2}{r}{\textbf{Category 3: Decentralization}} \\ [1ex] 
 \hline 
\citet{zhang2022sok} & SoK: Blockchain Decentralization \\[0.2ex] 
\hline
\citet{karakostas2022sok} & SoK: A Stratified Approach to Blockchain Decentralization\\[0.2ex]
\hline

\citet{raikwar2022sok} & SoK: Decentralized Randomness Beacon Protocols\\[0.2ex]
\hline
 
\end{tabular}
\caption{A Taxonomy of SoKs}
\label{tab1}
\end{table}

%% file: tabs/tab2.tex
\begin{table}[!htbp]
\fontsize{7pt}{\baselineskip}\selectfont
\setlength{\tabcolsep}{4pt}
\begin{tabular}{|l|r|}
 \hline \hline
Citation & Title \\ [1ex] 
 \hline \hline
\multicolumn{2}{r}{\textbf{Category 4: Applicability}} \\ [1ex] 
\hline 
\citet{gudgeon2020sok} &  SoK: Layer-Two Blockchain Protocols \\[0.2ex] 
\hline

\citet{bartoletti2021sok} &  SoK: lending pools in decentralized finance \\[0.2ex] 
\hline

\citet{werner2021sok} &  SoK: Decentralized Finance (DeFi) \\[0.2ex] 

\hline
\citet{xu2021sok} &  SoK: Decentralized Exchanges (DEX) with Automated Market Maker (AMM) Protocols \\[0.2ex] 
\hline

\citet{shi2021blockchain} & When Blockchain Meets Auction Models: A Survey, Some Applications, and Challenges \\[0.2ex] 
\hline

\citet{abuidris2019survey} & A survey of blockchain-based on e-voting systems \\[0.2ex] 
\hline

\citet{gadekallu2022blockchain} & Blockchain for the Metaverse: A Review \\[0.2ex] 
\hline

\citet{ali2019blockchain} & Blockchain and the future of the internet: A comprehensive review \\[0.2ex] 
\hline

\citet{cousaert2022sok} & Sok: Yield aggregators in defi \\[0.2ex] 
\hline

\citet{yu2022sok} & SoK: Play-to-Earn Projects \\[0.2ex] 
\hline

\citet{dotan2020sok} & SOK: cryptocurrency networking context, state-of-the-art, challenges \\[0.2ex] 
\hline

\citet{moin2020sok} & SoK: A classification framework for stablecoin designs \\[0.2ex] 
\hline

\citet{gan2021critical} & \thead{A critical review of blockchain applications to banking and finance:\\ a qualitative thematic analysis approach} \\[0.2ex] 
\hline

\citet{dasaklis2021sok} & Sok: Blockchain solutions for forensics \\[0.2ex] 
\hline

\citet{wang2021sok} & SoK: tokenization on blockchain \\[0.2ex] 
\hline

\citet{karantias2020sok} & Sok: A taxonomy of cryptocurrency wallets \\[0.2ex] 
\hline

\citet{clark2019sok} & SoK: demystifying stablecoins \\[0.2ex] 
\hline
\citet{jourenko2019sok} & SoK: A taxonomy for layer-2 scalability related protocols for cryptocurrencies \\[0.2ex] 
\hline

\citet{lande2018sok} & SoK: unraveling Bitcoin smart contracts \\[0.2ex] 
\hline
\citet{moin2020sok} & SoK: A Classification Framework for Stablecoin Designs \\[0.2ex] 
\hline

 \multicolumn{2}{r}{\textbf{Category 5: Governance and Regulations}} \\ [1ex] 
 \hline
 \citet{kiayias2022sok} & SoK: Blockchain Governance \\[0.2ex] 
\hline
 \citet{chatzigiannis2021sok} & SoK: Auditability and Accountability in Distributed Payment Systems \\[0.2ex] 
\hline
 \citet{kolachala2021sok} & SoK: Money Laundering in Cryptocurrencies \\[0.2ex] 
\hline

 \citet{casino2022sok} & SoK: Cross-border Criminal Investigations and Digital Evidence \\[0.2ex] 
\hline

 \citet{deuber2022sok} & \thead{SoK: Assumptions Underlying Cryptocurrency Deanonymizations\\--A Taxonomy for Scientific Experts and Legal Practitioners} \\[0.2ex] 
\hline


\multicolumn{2}{r}{\textbf{Category 6: System Design}} \\ [1ex] 
 \hline
 \citet{wang2020sok} & SoK: Diving into DAG-based blockchain systems \\[0.2ex] 
\hline
 \citet{bellaj2022sok} & SOK: a comprehensive survey on distributed ledger technologies \\[0.2ex]
\hline

\multicolumn{2}{r}{\textbf{Category 7: Cross-chain Interoperability}} \\ [1ex] 
 \hline
 \citet{zamyatin2021sok} & Sok: Communication across distributed ledgers \\[0.2ex] 
\hline
 \citet{wang2021sok} & Sok: Exploring blockchains interoperability \\[0.2ex] 
\hline
 \citet{eskandari2021sok} & Sok: Oracles from the ground truth to market manipulation \\[0.2ex] 
\hline
 \citet{lee2022sok} & SoK: Not Quite Water Under the Bridge: Review of Cross-Chain Bridge Hacks \\[0.2ex] 
\hline
 \citet{Rafael_survey_2022} & A Survey on Blockchain Interoperability: Past, Present, and Future Trends \\[0.2ex] 
\hline
 \citet{Rafael_solution_2022} & Do You Need a Distributed Ledger Technology Interoperability Solution? \\[0.2ex] 
\hline

\end{tabular}
\caption{The List of SoKs (continued)}
\label{tab2}
\end{table}

%% file: tabs/tab3.tex
\begin{table}[!htbp]
\caption{The Bigram of Blockchain Related SoK Titles (Top 25)}
\begin{tabular}{lr}
\toprule
                        bigram &  counts \\
\midrule
      (decentralized, finance) &       5 \\
         (distributed, ledger) &       3 \\
             (smart, contract) &       3 \\
          (blockchain, system) &       2 \\
          (stablecoin, design) &       2 \\
   (classification, framework) &       2 \\
          (ledger, technology) &       2 \\
       (framework, stablecoin) &       2 \\
(blockchain, decentralization) &       2 \\
               (finance, defi) &       2 \\
      (finance, decentralized) &       2 \\
             (blockchain, era) &       2 \\
       (dag-based, blockchain) &       1 \\
               (lending, pool) &       1 \\
             (ledger, lending) &       1 \\
         (across, distributed) &       1 \\
       (communication, across) &       1 \\
     (contract, communication) &       1 \\
         (confidential, smart) &       1 \\
  (tee-assisted, confidential) &       1 \\
        (system, tee-assisted) &       1 \\
       (solution, blockchains) &       1 \\
           (diving, dag-based) &       1 \\
              (client, diving) &       1 \\
               (light, client) &       1 \\
\bottomrule
\end{tabular}
\label{tab3}
\end{table}

%% file: tabs/tab4.tex
\begin{table}[!htbp]
\caption{The Bigram of Blockchain Related SoK Abstracts (Top 1-45)}
\begin{tabular}{lr}
\toprule
                        bigram &  counts \\
\midrule
      (blockchain, technology) &      25 \\
             (smart, contract) &      25 \\
          (blockchain, system) &      14 \\
            (future, research) &      13 \\
  (systematization, knowledge) &      12 \\
      (decentralized, finance) &      11 \\
         (research, direction) &      10 \\
               (lending, pool) &       8 \\
         (consensus, protocol) &       7 \\
               (finance, defi) &       7 \\
           (automated, market) &       6 \\
       (blockchain, metaverse) &       6 \\
                (recent, year) &       6 \\
               (market, maker) &       6 \\
      (cryptographic, concept) &       6 \\
           (security, privacy) &       6 \\
         (layer-two, protocol) &       6 \\
        (machine, replication) &       6 \\
              (state, machine) &       6 \\
                  (maker, amm) &       5 \\
(blockchain, interoperability) &       5 \\
          (digital, forensics) &       5 \\
              (yield, farming) &       5 \\
     (application, blockchain) &       5 \\
         (distributed, system) &       5 \\
              (open, research) &       5 \\
         (research, challenge) &       5 \\
         (distributed, ledger) &       5 \\
       (blockchain, ecosystem) &       5 \\
             (best, knowledge) &       5 \\
        (blockchain, security) &       4 \\
         (repeated, consensus) &       4 \\
             (finance, sector) &       4 \\
            (banking, finance) &       4 \\
             (transfer, asset) &       4 \\
                   (use, case) &       4 \\
              (privacy, issue) &       4 \\
           (future, direction) &       4 \\
  (decentralized, application) &       4 \\
  (cross-chain, communication) &       4 \\
              (body, research) &       4 \\
           (solution, finally) &       4 \\
      (off-chain, transaction) &       4 \\
           (financial, system) &       4 \\
             (open, challenge) &       4 \\
\bottomrule
\end{tabular}

\label{tab4}
\end{table}

%% file: tabs/tab5.tex
\begin{table}[!htbp]
\caption{The Bigram of Blockchain Related SoK Abstracts Continued (Top 46-90)}
\begin{tabular}{lr}
\toprule
                        bigram &  counts \\
\midrule
    (provide, systematization) &       4 \\
             (payment, system) &       4 \\
   (systematic, comprehensive) &       4 \\
              (bft, consensus) &       4 \\
             (protocol, along) &       4 \\
              (first, present) &       3 \\
         (security, guarantee) &       3 \\
             (finally, discus) &       3 \\
             (general, design) &       3 \\
              (payment, state) &       3 \\
            (work, blockchain) &       3 \\
         (layer-two, solution) &       3 \\
              (state, channel) &       3 \\
               (light, client) &       3 \\
                  (built, top) &       3 \\
                (third, party) &       3 \\
              (exchange, dexs) &       3 \\
          (existing, protocol) &       3 \\
            (system, security) &       3 \\
              (integral, part) &       3 \\
     (decentralized, exchange) &       3 \\
          (ledger, technology) &       3 \\
             (technology, dlt) &       3 \\
              (auction, model) &       3 \\
        (research, innovation) &       3 \\
          (existing, solution) &       3 \\
           (challenge, future) &       3 \\
            (service, however) &       3 \\
               (high, latency) &       3 \\
    (cryptographic, primitive) &       3 \\
            (existing, system) &       3 \\
         (potential, research) &       3 \\
        (solution, blockchain) &       3 \\
                  (two, major) &       3 \\
            (property, system) &       3 \\
              (provide, first) &       3 \\
             (scattered, body) &       3 \\
        (consensus, mechanism) &       3 \\
             (important, role) &       3 \\
         (application, domain) &       3 \\
              (paper, conduct) &       3 \\
         (conduct, systematic) &       3 \\
               (research, gap) &       3 \\
            (identify, future) &       3 \\
              (based, finding) &       3 \\
\bottomrule
\end{tabular}
\label{tab5}
\end{table}

%% file: tabs/tab6.tex
\begin{table}[!htbp]
\caption{The Top 28 blockchain project and cross-chain solutions: ranked by market value retrieved from coinmarketcap on Dec. 27, 2022 }
\begin{tabular}{rllrl}
\toprule
 Rank  &                   name & symbol &  Genesis &        Type \\
\midrule
     1 &               Bitcoin  &    BTC &     2009 &  Blockchain \\
     2 &               Ethereum &    ETH &     2015 &  Blockchain \\
     3 &    Binance smart chain &    BNB &     2017 &  Blockchain \\
     4 &    XRP Ledger (Ripple) &    XRP &     2021 &  Blockchain \\
     5 &                Cardano &    ADA &     2017 &  Blockchain \\
     6 &               Polkadot &    DOT &     2022 & Cross-chain \\
     7 &                   TRON &    TRX &     2017 &  Blockchain \\
     8 &                 Solana &    SOL &     2020 &  Blockchain \\
     9 &              Avalanche &   AVAX &     2020 &  Blockchain \\
    10 &              Chainlink &   LINK &     2017 & Cross-chain \\
    11 & The Open Network (TON) &    TON &     2018 &  Blockchain \\
    12 &                 Cosmos &   ATOM &     2016 & Cross-chain \\
    13 &                Stellar &    XLM &     2015 &  Blockchain \\
    14 &           Cronos Chain &    CRO &     2018 &  Blockchain \\
    15 &       Quant Overledger &    QNT &     2018 & Cross-chain \\
    16 &                Agorand &   ALGO &     2019 &  Blockchain \\
    17 &          NEAR Protocol &   NEAR &     2021 &  Blockchain \\
    18 &               Filecoin &    FIL &     2017 & Cross-chain \\
    19 &                 Hedera &   HBAR &     2019 &  Blockchain \\
    20 &      Internet Computer &    ICP &     2021 &  Blockchain \\
    21 &            EOS Network &    EOS &     2018 &  Blockchain \\
    22 &   MultiversX (Elrond)  &   EGLD &     2020 &  Blockchain \\
    23 &                   Flow &   FLOW &     2018 &  Blockchain \\
    24 &          Theta Network &  THETA &     2019 &  Blockchain \\
    25 &                  Tezos &    XTZ &     2018 &  Blockchain \\
    26 &                  Zcash &    ZEC &     2016 &  Blockchain \\
    27 &                 Klaytn &   KLAY &     2019 &  Blockchain \\
    28 &                   Dash &   DASH &     2014 &  Blockchain \\
\bottomrule
\end{tabular}
\label{tab6}
\end{table}

%% file: tabs/tab7.tex
\begin{table}[!htbp]
\caption{The Bigram of Blockchain Projects and Cross-chain solutions Titles (Top 10) }

\begin{tabular}{lr}
\toprule
                    bigram &  counts \\
\midrule
         (smart, contract) &       2 \\
      (public, blockchain) &       2 \\
    (blockchain, platform) &       2 \\
            (white, paper) &       2 \\
       (public, hashgraph) &       1 \\
         (ledger, stellar) &       1 \\
     (distributed, ledger) &       1 \\
    (network, distributed) &       1 \\
     (whitepaper, network) &       1 \\
      (cosmos, whitepaper) &       1 \\
         (network, cosmos) &       1 \\
           (open, network) &       1 \\
           (network, open) &       1 \\
         (oracle, network) &       1 \\
   (decentralized, oracle) &       1 \\
(evolution, decentralized) &       1 \\
         (step, evolution) &       1 \\
              (next, step) &       1 \\
         (chainlink, next) &       1 \\
      (dynamic, chainlink) &       1 \\
           (avax, dynamic) &       1 \\
             (token, avax) &       1 \\
           (native, token) &       1 \\
       (avalanche, native) &       1 \\
           (v0, avalanche) &       1 \\
\bottomrule
\end{tabular}
\label{tab7}
\end{table}

%% file: tabs/tab8.tex
\begin{table}[!htbp]
\caption{The Bigram of Blockchain Projects and Cross-chain Solutions Abstracts (Top 1-45)}
\begin{tabular}{lr}
\toprule
                           bigram &  counts \\
\midrule
                (smart, contract) &      27 \\
               (technical, paper) &       8 \\
                 (digital, asset) &       6 \\
           (consensus, algorithm) &       6 \\
            (blockchain, network) &       6 \\
     (decentralized, application) &       5 \\
               (virtual, machine) &       5 \\
             (public, blockchain) &       5 \\
                    (cosmos, hub) &       5 \\
           (byzantine, agreement) &       4 \\
     (decentralised, application) &       4 \\
       (blockchain, architecture) &       4 \\
            (transaction, second) &       4 \\
                  (beacon, chain) &       4 \\
                  (native, token) &       3 \\
          (development, platform) &       3 \\
                (payment, scheme) &       3 \\
               (storage, network) &       3 \\
                   (zone, cosmos) &       3 \\
       (distributed, application) &       3 \\
        (blockchain, application) &       3 \\
                      (end, user) &       3 \\
                   (proof, stake) &       3 \\
                (new, blockchain) &       3 \\
                (paper, proposes) &       3 \\
           (blockchain, platform) &       3 \\
          (programming, language) &       3 \\
         (blockchain, technology) &       3 \\
                    (paper, also) &       3 \\
                (introduces, new) &       2 \\
                (tendermint, bft) &       2 \\
                  (ibc, protocol) &       2 \\
             (communication, ibc) &       2 \\
(inter-blockchain, communication) &       2 \\
                      (hub, zone) &       2 \\
                      (zone, hub) &       2 \\
                   (well, suited) &       2 \\
                (cosmos, network) &       2 \\
               (address, problem) &       2 \\
          (network, architecture) &       2 \\
            (transaction, ledger) &       2 \\
        (transaction, throughput) &       2 \\
          (governance, mechanism) &       2 \\
                   (machine, evm) &       2 \\
                  (without, need) &       2 \\
\bottomrule
\end{tabular}
\label{tab8}
\end{table}

%% file: tabs/tab9.tex
\begin{table}[!htbp]
\caption{The Bigram of Blockchain Projects and Cross-chain Solutions Abstracts (Top 46-90)}
\begin{tabular}{lr}
\toprule
                           bigram &  counts \\
\midrule
                   (ba, protocol) &       2 \\
           (achieves, robustness) &       2 \\
                  (protocol, scp) &       2 \\
             (application, built) &       2 \\
               (set, transaction) &       2 \\
                      (next, set) &       2 \\
               (secure, scalable) &       2 \\
                   (algorand, us) &       2 \\
           (widespread, adoption) &       2 \\
             (network, providing) &       2 \\
            (application, across) &       2 \\
                   (broad, range) &       2 \\
             (ledger, technology) &       2 \\
            (distributed, ledger) &       2 \\
                (trusted, secure) &       2 \\
               (cronos, designed) &       2 \\
               (transaction, fee) &       2 \\
               (low, transaction) &       2 \\
           (million, transaction) &       2 \\
               (provides, useful) &       2 \\
                   (network, ton) &       2 \\
                   (used, encode) &       2 \\
                  (strong, focus) &       2 \\
                  (work, present) &       2 \\
           (network, communicate) &       2 \\
            (consensus, protocol) &       2 \\
              (handling, million) &       2 \\
                    (high, speed) &       2 \\
       (autonomous, organization) &       2 \\
        (functional, programming) &       2 \\
      (decentralized, autonomous) &       2 \\
                (important, part) &       2 \\
                     (cpu, power) &       2 \\
                 (longest, chain) &       2 \\
               (network, network) &       2 \\
          (peer-to-peer, network) &       2 \\
                   (open, source) &       2 \\
              (advanced, feature) &       2 \\
                  (open, network) &       2 \\
           (consensus, mechanism) &       2 \\
                  (hybrid, smart) &       2 \\
            (computing, resource) &       2 \\
           (off-chain, computing) &       2 \\
                (oracle, network) &       2 \\
               (new, blockchains) &       2 \\
\bottomrule
\end{tabular}
\label{tab9}
\end{table}